\documentclass[12pt,epsfig]{article}

\usepackage{graphicx}
\usepackage[usenames]{color}
\usepackage{epsfig}
\usepackage{latexsym}
\definecolor{rotblass}{cmyk}{0,0.07,0.07,0.001}

\newcommand\noi{\noindent}
\newcommand\beq{\begin{equation}}
\newcommand\eeq{\end{equation}}
\newcommand\beqn{\begin{eqnarray}}
\newcommand\eeqn{\end{eqnarray}}

\newcommand\bla{\color{black}}
\newcommand\gre{\color{green}}

\newcommand\bl{\color{blue}}
\newcommand\cy{\color{cyan}}
\newcommand\ma{\color{magenta}}
\newcommand\ye{\color{yellow}}
\newcommand\br{\color{RawSienna}}
\newcommand\red{\color{red}}
\definecolor{lg}{cmyk}{0,0,0,0.1}
\definecolor{lg2}{cmyk}{0,0,0,0.3}
\definecolor{la}{named}{Lavender}
\definecolor{lb}{named}{SkyBlue}
\definecolor{sg}{named}{SpringGreen}
\definecolor{do}{named}{BurntOrange}
\definecolor{or}{rgb}{1,0.8107,0.8077}

\title{
\begin{center}
{\Large\ma RESULTS  FROM SKM-200-GIBS ON MULTIPARTICLE}\\[0.2cm]
{\Large\ma AZIMUTHAL CORRELATIONS IN C-Ne AND C-Cu COLLISIONS AT
ENERGY OF 3.7 GeV PER NUCLEON}
\end{center}
\vspace{4cm}
}
\author{{\Large\bla L.Chkhaidze, T.Djobava,
L.Kharkhelauri}\\[0.3cm]
{\large\sl\bla High Energy Physics Institute, Tbilisi State
University,}\\[0.3cm]
{\large\sl\bla University St.9, 380086 Tbilisi, Georgia}\\[0.3cm]
{\large\sl\bla E-mail:djobava@sun20.hepi.edu.ge or Tamar.Djobava@cern.ch}}
\begin{document}

\maketitle

\newpage
\topmargin 0cm
\oddsidemargin -0.4cm
%\textheight=24.cm
%\textwidth=16.cm

\hspace{12.cm} {\bla TSU-HEPI-2000-5} \\

\centerline{
{\Large\ma ABSTRACT}}
\vspace{1.cm}
%\large\bla
\noi
{\color{black}\Huge$\bullet$}
The transverse momentum technique is used to analyse  charged-particle
exclusive data in the central {\ma C-Ne} and {\ma C-Cu} interactions at energy of
{\ma 3.7 GeV} per nucleon.
 Clear evidence of {\ma in-plane} and {\ma out-of-plane} (squeeze-out)
flow effects for
protons and $\pi^{-}$ mesons have been obtained.
In  C-Ne interactions in-plane flow of $\pi^{-}$ mesons is in the same
direction as for the protons, while in  C-Cu collisions pions show
antiflow behaviour.
From the transverse momentum and azimuthal distributions of
 protons and $\pi^{-}$ mesons with respect
to the reaction plane, the  flow $F$ (the measure of the collective 
transverse momentum transfer in the reaction plane) and
the parameter $a_{2}$ (the measure of the anisotropic
emission strength)
have been extracted. The flow effects increase with the mass of the particle
and the mass number of target $A_{T}$.\\[0.8cm]
%{\color{PineGreen}\Huge$\bullet$}
{\color{black}\Huge$\bullet$}
The comparison of our in-plane flow results with flow data for various
projectile/target configurations was made using the
{\ma scaled flow
$F_{S}$=$F/(A_{P}^{1/3}+A_{T}^{1/3})$}. $F_{S}$ demonstrates a common scaling
behaviour for flow values from different systems.\\[0.8cm]
{\color{black}\Huge$\bullet$}
{\ma The Quark Gluon String Model (QGSM)} was used for the comparison with
the experimental data.
The QGSM yields a signature of in-plane and out-of-plane flow effects
in C-Ne and C-Cu collisions for protons.

\newpage
\topmargin 0cm
\oddsidemargin -0.4cm

\vspace{1cm}
%\large\bla
\noi
\par
One of the {\ma main goals} of relativistic heavy-ion collision
experiments is to study nuclear matter under extreme conditions of high
densities and temperatures, i.e. to learn more about the nuclear equation
of state {\ma (EOS)}.
An increasing number of observables which are accessible through
heavy-ion collisions has been found to be sensitive to the EOS.
In order to study the EOS, collective effects, such as
the directed transverse flow (the bounce-off of cold
spectator matter in the reaction plane) {\bl [1]}
and the elliptic flow
(the squeeze-out of hot and compressed participant matter perpendicular
to the reaction plane) {\bl [2]}, are frequently used.
    According to theoretical
    calculations, flow studies can provide information  on the
    collision dynamics as well as on
    a possible phase transition to soft quark matter.
\par

    {\ma Collective flow} is the consequence of the pressure
    build-up in the high density zone through the short range repulsion
    between nucleons, i.e. through compressional energy. This effect
    leads to characteristic, azimuthally asymmetric sidewards emission
    of the reaction products.
     While the transverse flow
    in the reaction plane is influenced by the cold matter
    deflected by the overlap region of the colliding nuclei, the squeeze-
    out is caused by the hot and compressed matter from the interaction
    region which preferentially escapes in the direction perpendicular
    to the reaction plane, unhindered by the presence of the projectile
    and target spectators.
\par

The efforts to determine the EOS and the more general aspect of producing
high densities over extended regions have led to a series of
experiments studying relativistic nucleus-nucleus collisions
at BEVALAC (Berkeley), GSI-SIS (Darmstadt), JINR (Dubna),
AGS (Brookhaven National Laboratory) and SPS (CERN).
\par

Using the transverse momentum technique developed by
{\sl\color{magenta}
P.Danielewicz and G.Odyniec} {\bl [3]},
nuclear collective flow has already been observed
for protons, light nuclei, pions and $\Lambda$ - hyperons emitted in
nucleus-nucleus
collisions at the energies of 0.4$\div$1.8 GeV/nucleon of BEVALAC, GSI-SIS
{\bl[4-12]},
 at 11$\div$14 GeV/nucleon of AGS {\bl [13,14]} and at 158 GeV/nucleon
of CERN {\bl [15]}. The discovery
of collective sidewards flow in Au+Au at the AGS was a major
highlight of 1995 {\bl [14]}.
\par
 We present experimental results obtained from the in
 and out-plane
transverse momentum analysis {\ma for protons and $\pi^{-}$ mesons}
in central C-Ne and C-Cu
interactions at E=3.7 GeV per nucleon.
Data were obtained on
the {\ma SKM-GIBS} set-up at JINR. The obtained signature
 shows the persistence of collective flow phenomena
all the way up to AGS energies. Our results,
obtained by streamer chamber technique,
provide quantitative information on
the transverse and out-of-plane  (squeeze-out) elliptic flows
and their dependence
on beam energy and projectile/target mass complementing the experimental
data available from BEVALAC, GSI-SIS and AGS.
\par
 {\ma SKM-GIBS} consists of a 2 m streamer chamber, placed in a magnetic field
of 0.8 T, and a triggering system. The streamer chamber was exposed to the
beam of C nuclei accelerated in the synchrophasotron
up to the energy of 3.7 GeV/nucleon. The thickness of the solid target in
 the shape  of a thin Cu disc was 0.2 g/cm$^{2}$.
 Neon gas filling the chamber
also served as a nuclear target. The experimental set-up and the logic of
the triggering system are presented in {\bl Fig.1}
 The triggering system  allowed  the  selection  of
{\ma "inelastic"}  and {\ma "central"} collisions.
\par
    The inelastic trigger was selecting all inelastic  interactions of
incident nuclei on a target.
\par
    The central trigger was selecting events with no charged projectile
spectator fragments (with  $P/Z>3$ GeV/c ) within a cone of  half-angle
$\Theta_{ch}$ = 2.4$^{0}$ or  2.9$^{0}$.   The trigger efficiency
for events  with  a single charged particle in the cone was 99$\%$
The biases
and correction procedures were discussed in detail in Refs. {\bl [16,17]}. The
ratio  $\sigma_{cent}$/$\sigma_{inel}$  that characterizes the
centrality of selected events is  - (9$\pm$1)$\%$ for C-Ne
and (21$\pm$3)$\%$ - for C-Cu.
In {\bl Table 1} the number of events is presented. Average
measurement errors of the momentum and production angles are:
for protons $<\Delta P/P>$= (8$\div$10)$\%$, $\Delta$$\Theta$ =1$^{0}$$\div$2$^{0}$;
for pions  ~~~$<\Delta P/P>$= 5$\%$, $\Delta$$\Theta$ =0.5$^{0}$.
\par
Data have been analysed {\ma event by event using the transverse momentum
technique of
P.Danielewicz and G.Odyniec} {\bl [3]}.
Using this method, nuclear {\ma collective flow}
for protons has been observed
in central C-Ne and C-Cu interactions at
a momentum of P=4.5 GeV/c/N (E=3.7 GeV/nucl) with the SKM-200 set-up at JINR
and described in detail elsewhere {\bl [18]}.
The results for protons in C-Cu interactions presented here
are obtained on a statistics twice as large as in {\bl [18]}.
 Results for both C-Ne and
C-Cu collisions are presented in terms of the normalized rapidity
in the laboratory system $y/y_{p}$ where $y_{p}$=2.28 is the
 projectile rapidity.
 P.Danielewicz and G.Odyniec  have proposed to present the data
 in terms of the mean transverse momentum
per nucleon in the reaction plane $<P_{x}(Y)>$ as a function of rapidity.
The vector
\begin{center}
%\centerline{
{\ma
$\overrightarrow{Q_{j}}=\sum\limits_{i\not=j}\limits^{n} \omega_{i}
\overrightarrow{P_{{\perp}i}}$}
\end{center}
was used fo  the reaction plan for each event
The reaction plane is the plane containing
$\overrightarrow{Q_{j}}$
and the beam axis,
where $P_{{\perp}i}$ is the transverse momentum  of particle $i$, and $n$ is
the number of particles in the event.
\noi
 Pions were not included.
The weight
{\ma $\omega_{i}$= $y_{i}$ - $<y>$} {\bl [9]}, where  $<y>$ is the
average rapidity of the participant protons, calculated for each event.
The participant protons are those which are neither projectile nor
target fragments.
Average multiplicities of analysed  protons $<N_{p}>$
 are listed in {\bl Table 1}.
\par
 The transverse momentum of each
particle  in the  estimated reaction plane is calculated as
\begin{center}
%\centerline{
{\ma
$P_{xj}\hspace{0.01cm}^{\prime} = \{ \overrightarrow{{Q_{j}}}\cdot
\overrightarrow{P_{{\perp}j}} /
\vert\overrightarrow{Q_{j}}\vert\}$}
\end{center}
\par
%\noi
 The average transverse momentum
$<P_{x}\hspace{0.01cm}^{\prime}(Y)>$
%$<P_{x}(Y)>$
is obtained by averaging over all
events in the corresponding intervals of rapidity.
\par
%\noi
For the event by event analysis it is
necessary to perform an identification
of $\pi^{+}$ mesons, the admixture of which
amongst the charged positive particles is about {\ma (25$\div$27)$\%$} . The
identification has been carried out on the statistical basis using
 the two-dimentional ( $P_{\parallel}$, $P_{\perp}$ ) distribution.
It was assumed, that  $\pi^{-}$  and $\pi^{+}$ mesons
hit a given cell in the plane
( $P_{\parallel}$, $P_{\perp}$ ) with equal probability.
 The difference in multiplicities of $\pi^{+}$  and $\pi^{-}$
in each event was required to be no more than 2. After
this procedure the admixture of $\pi^{+}$ did not exceed
{\ma (5$\div$7)$\%$}.
 The temperature of the identified protons agrees with our
previous result {\bl [19]}, obtained by the method of spectra subtraction.
\par
%\noi
It is known {\bl [4]} that the estimated reaction plane
differs from the true  one,
due to the finite number of particles in each event.
The component $ P_{x}$ in the true reaction plane is systematically larger
then the component $P_{x}\hspace{0.01cm}^{\prime}$
in the estimated plane, hence
{\ma $<P_{x}>=<P_{x}\hspace{0.01cm}^{\prime}>/<cos\varphi> $},
where $\varphi$ is the angle between the estimated and true planes.
\par
%\noi
The  correction factor
{\ma $K$=1 $/$ $< cos\varphi >$} is subject to a large uncertainty,
especially for low multiplicity.
In order to determine  $< cos\varphi>$ according to {\ma [3]}, we
divided each event randomly into two equal sub-events and calculated vectors
{\ma $\overrightarrow{Q}$}
for each of these sub-events. Then $\varphi$ was estimated  as the angle
between these two vectors.
The values of $K$
averaged over all multiplicities are: $K=1.27\pm0.08$
for C-Ne,  $K=1.31\pm0.04$  for C-Cu.
\par
%\noi
{\bl Figs.2,3}  show the dependence of $<P_{x}>$
on the normalized rapidity $y/y_{p}$ in the laboratory system
for protons and pions
in C-Ne {\bl (Fig.2)} and C-Cu {\bl (Fig.3)}
collisions. For protons the data points are multiplied by the factor
 $K$ described above to correct
for the deviation from the true reaction plane.
The data exhibit the typical $S$-shape behaviour
which demonstrates the collective transverse momentum transfer between the
forward and backward hemispheres.
\par
%\noi
Based on the mean transverse momentum distributions we can extract  an
observable --
the transverse flow
{\ma $F=<P_{x}>/d(y/y_{p})$}, the slope
of the momentum distribution at midrapidity. It is a measure of the amount
of collective transverse
momentum transfer in the reaction. Technically, $F$ is obtained by fitting
the central part of the dependence of
$<P_{x}>$ on $y/y_{p}$
with a linear function , with  the slope  equal to the flow $F$.
The fit was done for $y/y_{p}$ between 0.01 $\div$ 0.90.
The straight lines in {\bl Figs. 2,3} show the results of this fit.
The values of $F$ are listed in {\bl Table 1}.
We have analysed the influence of the admixture
of ambiguously identified $\pi^{+}$ mesons  on the results, and the quoted
 error in the flow $F$ includes both the statistical and systematical
 uncertaintes.
One can see from the {\bl Table 1}, that  with the increase of the mass
number of
the target A$_{T}$ the value of $F$ increases. A similar
tendency was observed at lower energies {\bl [4-7,10]}.
\par
%\noi
It is of great  interest to compare the flow values for a wide range of data.
A way of comparing the energy dependence of flow values for different
projectile/target mass combinations was suggested by
%\par
%\noi
{\sl\ma
A.Lang et al} {\bl [20]} and
first used by {\sl\ma J.Chance} in {\bl [10]}. To allow for
different projectile/target ($A_{P}$,$A_{T}$)  mass systems, they divided
the flow values by ($A_{P}^{1/3}+A_{T}^{1/3}$) and called
{\ma
$F_{S}=F/(A_{P}^{1/3}+A_{T}^{1/3})$}  the scaled flow.
\par
%\noi
{\bl Fig.4} shows a plot of $F_{S}$ versus the projectile energy per
nucleon.
Fig.4 presents our results and the data from the EOS
{\bl [10,21]}, E-895, E-877 {\bl [21]},
FOPI {\bl [12]} experiments along with the values derived from the
Plastic Ball {\bl [7,11]} and
the Streamer Chamber {\bl [4,9]} experiments for a variety of energies and
mass
combinations.
The values of flow $F$ for E-895, E-877 are taken from {\bl Fig.5} of
{\bl [21]} and recalculated in terms of $F_{S}$.
For the EOS and the Plastic Ball data all the isotopes of
$Z$=1 and 2 are included, except for {\bl [11]} where the data are given
only
for $Z$=1.
 The Streamer Chamber data {\bl [4,9]} normally include
all protons, whether free or bound in clusters as in our case.
 The scaled flow
$F_{S}$ follows, within the uncertainties, a common trend with an initial
steep rise and then a gradual decrease.
It is worth noting
that  the data obtained by streamer chamber technique (including our results)
 are  slightly higher than those obtained by the
electronic experiments.
This may be caused by a small mixture of bound protons (deutons, $^{3}H$,
$^{4}He$).
\par
%\noi
Several theoretical models of nucleus-nucleus collisions at
high energy have been
proposed {\bl [22]}. {\ma The Quark Gluon String Model
(QGSM) } {\bl[23]} was used for a comparison with experimental data. The QGSM is based
on the Regge and string phenomenology of particle production in inelastic
binary hadron collisions {\bl [24]}.
The QGSM simplifies the nuclear effects (neglects the potential
interactions between hadrons, coalescence of nucleons and etc.).
A detailed description and
comparison of the QGSM with experimental data in a wide energy range can be
found in paper {\bl [25]}.
 The model
yields a generally good overall fit to most experimental data {\bl [25]}.
\par
We have generated C-Ne and C-Cu interactions using Monte-Carlo generator
{\ma COLLI}, based on the QGSM and then
traced through the detector and trigger filter..
\par
In the generator COLLI there are two possibility to generate events:
1) at not fixed impact parameter $b$ and 2)at fixed $b$.
From the b distributions we obtained the mean values {\ma $<b>$ =2.20 fm for C-Ne}
collisions and {\ma $<b>$=2.75 fm for C-Cu} and total samples of events for these
$<b>$ had been generated ({\bl Table.1}).
 The QGSM overestimates the production of low
momentum protons  with $P<$0.2 GeV/c, which  are mainly the target fragments
and were excluded from the analysis.  From the analysis of generated events
the protons with deep angles greater  60$^{0}$ had been excluded, because
 such vertical tracks  are registered with less efficiency on the experiment.
The QGSM yields a significant flow signature for protons, which follows trends similar
to the experimental data {\bl Figs 2,3}.
 The values of $F$, obtained
from the QGSM for protons are listed in {\bl Table 1}. One can see, that the QGSM
slightly overestimates the flow for C-Ne and underestimates for C-Cu.
 This model underestimates also the
transverse flow at BEVALAC energies (1.8 GeV/n) {\bl [3,4]}. As shown by H.Stocker
and  Greiner {\bl [26]}, the reason that the  QGSM fails to reproduce the flow
data in the
energy region of 1$\div$5 GeV/n, is the negligence of mean-field effects.
\par
%\noi
 In view of
 the strong coupling between the nucleon and pion, it is interesting to
know if pions also have a collective flow behaviour and how the pion flow
 is related to the nucleon flow.
\par
For this purpose the reaction plane was defined for the participant protons,
and the transverse momentum of each $\pi^{-}$ meson have been  projected onto
this reaction plane.
{\bl Figs. 2,3} show the dependence of $<P_{x}>$
on the normalized rapidity $y/y_{p}$ in the laboratory system
for $\pi^{-}$ mesons
in C-Ne and C-Cu collisions.
The data exhibit the typical  $S$-shape behaviour
as for the protons.
The values of flow $F$ for $\pi^{-}$ mesons
are: for C-Ne collisions {\ma $F=29 \pm 5$ MeV};
for C-Cu --- {\ma $F=-47 \pm 6$ MeV}.
The straight lines in {\bl  Fig. 2,3} show the results of this fit.
The fit was done in the following intervals of $y/y_{p}$: 0.04 $\div$ 0.7
for C-Ne ;  -0.06 $\div$ 0.6  for C-Cu .
The absolute value of $F$ increases
with the mass number of target $A_{T}$, indicating the rise of the
collective flow effect. The similar tendency was observed in [8]
for $\pi^{-}$ and $\pi^{+}$ mesons in Ne-Naf, Ne-Nb and Ne-Pb interactions
at 800 MeV/nucleon energy.
\par
%\noi
 One can see from {\bl Fig.2,3}, that for C-Ne collisions the
 $ \overrightarrow {P_{x}}$
for pions is directed in the same direction as for protons i.e.
flows of protons and pions are correlated, while for C-Cu interactions the
$ \overrightarrow{P_{x}}$ of $\pi^{-}$ mesons is directed oppositely to that of
the protons (antiflow).
\par
\noi
As obtained in Ref. {\bl [27]},
at AGS energy of 11 GeV/nucleon the flow
of $\pi^{+}$ mesons is in the direction
opposite to the protons, similarly to the observations
 in semi-central
Pb-Pb collisions at energy 158 GeV/nucleon in WA98 collaboration at SPS CERN
{\bl [15]}. The magnitude of the directed flow in {\bl [15]} is found to
be
significantly smaller than that observed at AGS energies.  Thus, it seems that
the flow effects for pions decrease with increased energy.
Theoretical calculations in the framework of
the Isospin Quantum Molecular Dynamics (IQMD) model
have predicted {\bl [28]} the existence of pion antiflow at projectile-
and
target rapidities
for Au-Au collisions at GSI-SIS  energies 1 GeV/nucleon. On the other hand
within the framework of the relativistic transport model (ART 1.0)
{\bl [29]} for
heavy-ion collisions (Au-Au) at AGS energies, pions are found to have a weak
flow behaviour.
\par
%\noi
The origin of the particular
shape of the $\overrightarrow{P_{x}}$ spectra for pions was studied
in {\bl [28-31]}.
The investigation revealed that
the origin of
the  in-plane transverse momentum of pions is
the pion scattering process (multiple $\pi N $ scattering) {\bl [28]}
and the pion absorption {\bl [30,31]}.  However, in {\bl [29]} it was
found that
 pions show a weak flow behaviour in central collisions  due to the
flow of baryon resonances from which they are produced.
\par
%\noi
The anticorrelation of nucleons and pions was explained
in {\bl [28]} as
due
to multiple $\pi ~N$ scattering. However, in {\bl [29,31]} it was shown
that
 anticorrelation is a manifestion of the nuclear shadowing effect
of the target- and projectile-spectators  through both pion rescattering and
reabsorptions.  In our opinion, our results indicate, that the flow behaviour
of $\pi^{-}$ mesons in light system C-Ne is due to the flow of
$\Delta$ resonances, whereas the antiflow behaviour in C-Cu collisions is the
result of the nuclear shadowing effect.
\par
\noi
The preferential emission of particles in the direction
perpendicular to the
reaction plane (i.e. "squeeze-out")
is particularly interesting since it is the only way the nuclear
matter might escape without being rescattered by spectator remnants of
the projectile and target, and is expected to provide direct information
on the hot and dense participant region formed in high energy
nucleus-nucleus interactions. This phenomenon, predicted by hydrodynamical
calculations {\bl [2]}, was clearly identified in the experiments
{\bl [32]}
by observation of an enhanced out-of-plane emission of
protons, mesons and charged fragments.
%For beam energies of 1 $\div$ 11
%GeV/nucleon the elliptic flow is the result of a strong competition between
%the early  "squeeze-out" and the late stage  "in-plane flow"
%{\bl [33]}. The magnitude
%and the sign of elliptic flow depend on two factors: a) the pressure built up
%in the compression stage compared to the energy density and b) the passage
%time of the projectile and target spectators.
\par
%\noi
In order to extend these
investigations, we have studied the azimuthal angular distributions
\noi
($\phi$)
of the pions and protons.
The angle $\phi$ is the angle of the transverse momentum of each particle
in an event with respect to the
reaction plane  ({\ma $cos\phi=P_{x}/P_{t}$}).
The analysis was restricted only to the mid-rapidity region by applying
a cut around the center of mass rapidity.
{\bl Figs. 5,6} show
distributions for protons and $\pi^{-}$ mesons in C-Ne {\bl Fig.5}
and C-Cu collisions {\bl Fig.6}.
For visual presentation the data on C-Cu were shifted upwards.
For $\pi^{-}$ mesons the analysis was performed from 0 to 180$^{0}$ due to
lower statistics than for protons. The azimuthal angular distributions
for the protons and pions
show a maxima at $\phi$=90$^{0}$  and
$270^{0}$ with respect to the event plane.
This maxima are associated with  preferential particle emission perpendicular to
the reaction plane (squeeze-out, or elliptic flow).  Thus a clear signature of an
out-of plane signal is
evidenced.
\par
%\noi
To treat the data in a quantitative way the azimuthal distributions were
fitted  by polynomial:
{\ma $dN/d\phi=a_{0}(1+a_{1}cos\phi+a_{2}cos2\phi)$}
%\par
\noi
 The anisotropy factor $a_{2}$ is negative for out-of-plane enhancement
(squeeze-out) and is the measure of the strength of the anisotropic
emission.
The values of the coefficient $a_{2}$  extracted from the
azimuthal distributions of protons and $\pi^{-}$ mesons are presented
in {\bl Table 2}.
 The fitted curves are superimposed
on the experimental distributions ({\bl Figs.5,6}). The QGSM data for protons
in C-Ne and C-Cu collisions are also superimposed in {\bl Fig.5} and the values
of $a_{2}$ are listed in {\bl Table 2}. One can see, that the model describes
the experimental azimuthal distributions.
\par
%\noi
The values of $a_{2}$ are used to quantify the ratio $R$
of the number of particles
emitted in the perpendicular direction to the number of particles emitted in
the reaction plane,
which represents the magnitude of the out-of-plane emission signal:
{\ma  $R=(1-a_{2})/(1+a_{2})$}.
A ratio $R$ larger than unity implies a preferred
out-of-plane emission. The values of $R$ are listed in {\bl Table 2}.
One can see that  $a_{2}$ and $R$ are increasing, both
for protons
and $\pi^{-}$ mesons, with increasing the transverse momentum
and the mass number of target $A_{T}$ and also with
narrowing of the cut applied around the center of mass rapidity.
The squeeze-out effect
is more pronounced for protons than for $\pi^{-}$ mesons.
Our results on rapidity, mass and transverse momentun
dependence of the azimuthal anisotropy are consistent with analyses from
Plastic Ball, FOPI, Kaos, TAPS {\bl [32,33]} collaborations and
are confirmed by IQMD calculations {\bl [34]}.
%\par
%\noi
In experiments (E-895, E-877, EOS) {\bl [36]} at AGS and SPS (CERN) (NA49)
energies the elliptic flow is typically studied at midrapidity and quantified
in terms of the second  Fourier coefficient
{\ma  $v_{2}=<cos2\phi>$}.
The Fourier coefficient $v_{2}$ is related to $a_{2}$ via
 $v_{2} \approx a_{2}/2$.
We have estimated $v_{2}$ both for C-Ne and C-Cu. The
dependence of the  elliptic flow
excitation function (for protons) on energy E$_{lab}$ is
displayed in {\bl Fig.7}.
 Recent calculations have made specific predictions for the
beam energy dependence of elliptic flow for Au-Au collisions at 1 $\div$ 11
GeV/nucleon {\bl [33]}. They indicate a transition from negative to
positive elliptic
flow at a beam energy {\ma E$_{tr}$}, which has a marked sensitivity to
the
stiffness of the EOS. In addition, they suggest that a phase transition to the
Quark-Gluon Plasma (QGP) should give a characteristic signature in the elliptic
flow excitation function due to the significant softening of the EOS. One can
see from {\bl Fig.7}, that the excitation function $v_{2}$ clearly shows
an
evolution from negative to positive elliptic flow within the region 2 $\leq
E_{beam} \leq $ 8 GeV/nucleon and point to  an apparent
transition energy
E$_{tr} \sim $ 4 GeV/nucleon.
\newpage
%\noi
\begin{center}
{\ma   SUMMARY}
\end{center}
\par
{\color{black}\Huge$\bullet$}
 We have reported experimental results,
presented
in terms of the mean transverse momentum per nucleon projected onto the
 reaction plane,
as a function of
the  normalized rapidity $y/y_{p}$ in laboratory system.
\par
{\color{black}\Huge$\bullet$}
We have determined
the flow {\ma $F$}, defined as the slope
at midrapidity.  $F$ increases
with the mass of the particle and mass number of target $A_{T}$.
\par
{\color{black}\Huge$\bullet$}
In C-Ne interactions flow of $\pi^{-}$ mesons is in the {\ma same} direction
as that of the protons, while in C-Cu collisions
pions show  the {\ma antiflow} behaviour.
The comparison of our in-plane flow results with flow data for various
projectile/target configurations was made using the scaled flow
{\ma
$F_{S}$=$F/(A_{P}^{1/3}+A_{T}^{1/3})$}.
$F_{S}$ demonstrates a common scaling
behaviour for flow values from different systems.
\par
{\color{black}\Huge$\bullet$}
A clear signature of an out-of-plane flow (squeeze-out)
have been obtained
from the azimuthal distributions of protons and $\pi^{-}$ mesons with respect
to the reaction plane at mid-rapidity region.
\par
{\color{black}\Huge$\bullet$}
Azimuthal distributions have been parametrized
by a second order polynomial function, and the parameter $a_{2}$ of the
anisotropy term $a_{2}cos2\phi$ have been extracted.
The ratio
{\ma  $R=(1-a_{2})/(1+a_{2})$}
was also calculated. The squeeze-out effect was shown to
increase with the transverse momentum,
mass number of target {\ma $A_{T}$} and also with narrowing of the rapidity
range. It is more pronounced
for protons than for $\pi^{-}$ mesons.
\par
{\color{black}\Huge$\bullet$}
{\ma The Quark Gluon String Model (QGSM)} was used for the comparison with
the experimental data.
The QGSM yields a signature of in-plane and out-of-plane flow effects
in C-Ne and C-Cu collisions for protons.

%\    \par
%\newpage
\vspace{2.cm}
%\centerline{
\begin{center}
{\ma  ACKNOWLEDGEMENTS }
\end{center}
%\par
%\   \par
\noi
 We would like to thank
{\ma
M.Anikina, A.Golokhvastov, S.Khorozov and} \\ {\ma J.Lukstins}
for  fruitful collaboration during the obtaining of the data.
We are very grateful to
{\ma Z. Menteshashvili and V.Kartvelishvili} for reading the manuscript
and many valuable remarks.

\pagebreak
%\par
\noi

\newpage
{\color{blue} Table 1}. The number of experimental
events,  the average  multiplicity of participant
protons $<N_{p}>$,
 the correction factor $K$ and
 the flow  $F$ for protons and $\pi^{-}$ mesons.
\   \par
\    \par
\     \par
%\noi
%\colorbox{SpringGreen}{\ma
\begin{tabular}{|l|c|c|}    \hline
&     &   \\
\hspace{0.5cm} &\hspace{1.cm} {\ma C-Ne} \hspace{1.3cm} &\hspace{1.3cm}
{\ma C-Cu}\hspace{1.6cm}        \\
&     &   \\
\hline
  Number of exper. events  &   {\bl 723}     &   {\bl 667}     \\
\hline
  Number of generated events  &   {\bl 8400}     &   {\bl 9327}     \\
			      &  {\bl $<b>$=2.20}&  {\bl $<b>$=2.75} \\
\hline
   $<N_{p}>$        &  {\bl 12.4 $\pm$ 0.5}   &  {\bl 19.5$\pm$ 0.6}   \\
\hline
  $K$=1/$< cos\varphi>$  &  {\bl 1.27 $\pm$ 0.08} &  {\bl 1.31 $\pm$ 0.04}
\\
\hline
  $F_{exp}$  for protons (MeV/c)   &  {\bl 134  $\pm$12}  & {\bl 198$\pm$13}     \\
\hline
  $F_{mod}$  for protons (MeV/c)   &  {\bl 160  $\pm$10}  & {\bl 190$\pm$9}     \\
\hline
  $F_{exp}$ for $\pi^{-}$ mesons (MeV/c) &   {\bl 29 $\pm$5}    & {\bl -47$\pm$6}
\\
\hline
\end{tabular}
%}
\newpage
{\color{blue} Table 2}.
The values of the parameter $a_{2}$ and the ratio $R$
for protons and $\pi^{-}$ mesons extracted
from the azimuthal distributions  fitted by\\
{\ma $dN/d\phi=a_{0}(1+a_{1}cos\phi+a_{2}cos2\phi)$}.
\   \par
\    \par
\     \par
%\noi
%\colorbox{SpringGreen}{\ma
\begin{tabular}{|c|c|c|c|c|}    \hline
&  &  &  &  \\
$A_{p}-A_{T}$ &Particle  &Applied Cut   & {\ma $ a_{2}$} &{\ma  $R$} \\
&  &  &  &  \\
\hline
&Protons&$-1\leq y_{cm}\leq1$& {\bl -0.049$\pm$0.014}& {\bl 1.10$\pm$0.03} \\
\cline{3-5}
&  & $-1\leq y_{cm}\leq1$; $P_{T}\geq 0.3$
GeV/c&{\bl -0.074$\pm$0.014} &{\bl 1.16$\pm$0.04} \\
\cline{2-5}
{\ma C-Ne} & $\pi^{-}$ mesons& $-1\leq
y_{cm}\leq1$&{\bl -0.035$\pm$0.013} &{\bl 1.07$\pm$0.04} \\
\cline{3-5}
&   & $-1\leq y_{cm}\leq1$; $P_{T}\geq 0.1$ GeV/c &
{\bl -0.050$\pm$0.014}& {\bl 1.09$\pm$0.03}\\
\hline
&Protons (mod.)  & $-1\leq y_{cm}\leq1$&{\bl -0.058$\pm$0.004}&{\bl 1.12$\pm$0.01} \\
\cline{2-5}
%\hline
& & $-1\leq y_{cm}\leq1$&{\bl -0.065$\pm$0.014}&{\bl 1.14$\pm$0.04} \\
\cline{3-5}
&Protons (exp.)   & $-1\leq y_{cm}\leq1$; $P_{T}\geq 0.3$&
{\bl -0.081$\pm$0.014}& {\bl 1.18$\pm$0.05} \\
\cline{3-5}
{\ma C-Cu}&  & $-0.6\leq y_{cm}\leq0.6$&{\bl -0.077$\pm$0.017}&
{\bl 1.17$\pm$0.04}\\
\cline{3-5}
& & $-0.6\leq y_{cm}\leq0.6$; $P_{T}\geq 0.3$&{\bl
-0.088$\pm$0.020}&{\bl 1.19$\pm$0.06} \\
\cline{2-5}
& $\pi^{-}$ mesons& $-1\leq y_{cm}\leq1$&{\bl
-0.041$\pm$0.013}&{\bl 1.08$\pm$0.03} \\
\cline{3-5}
&   & $-1\leq y_{cm}\leq1$; $P_{T}\geq 0.1$ GeV/c &{\bl
-0.056$\pm$0.015}&{\bl 1.12$\pm$0.04}\\
\hline
\end{tabular}
%}
%\end{document}
\pagebreak
%--------------------fig.1------------------------
\begin{figure}
\begin{center}
\epsfig{file=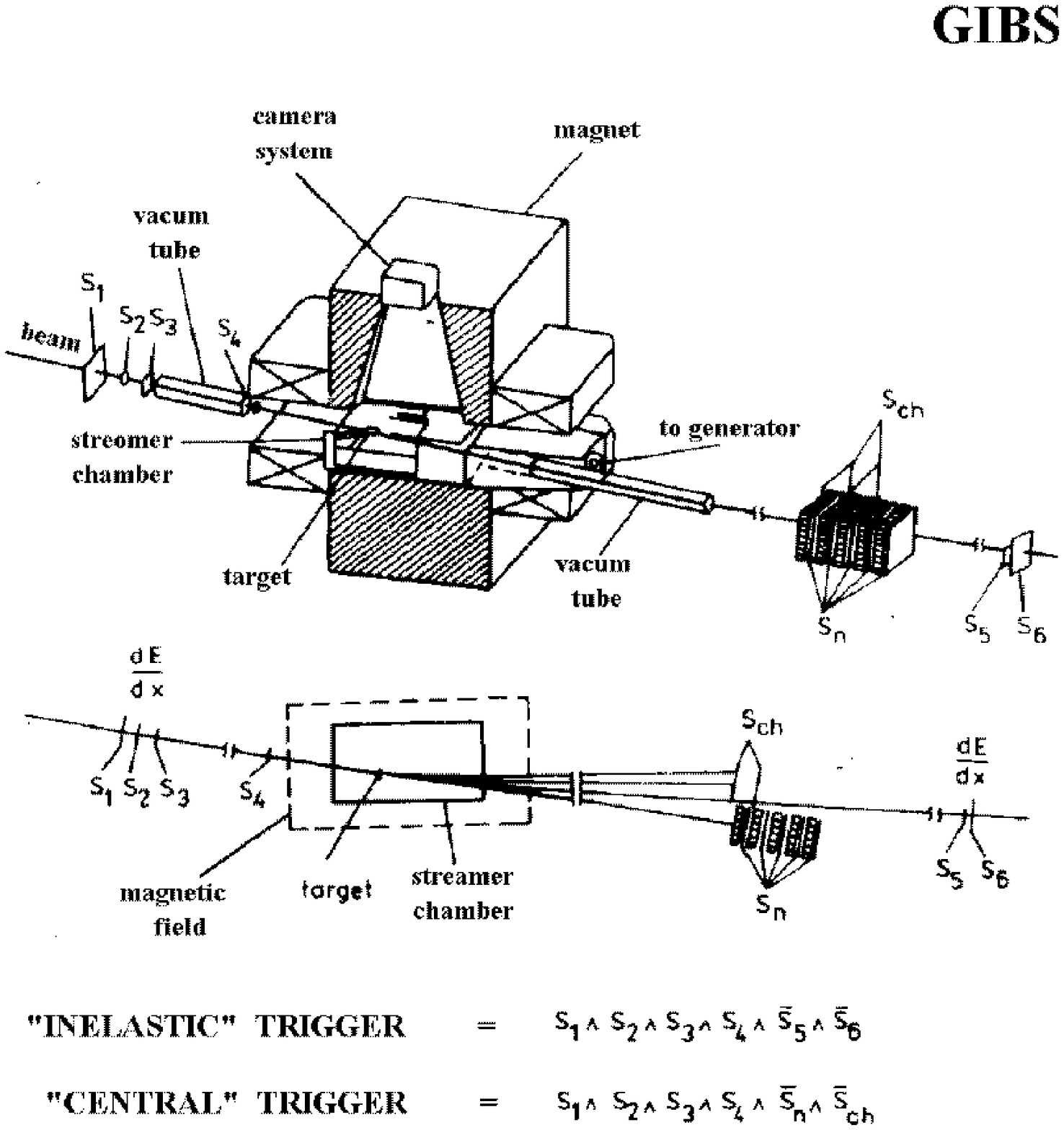,bbllx=0pt,bblly=0pt,bburx=594pt,bbury=842pt,
width=18cm,angle=0}
\end{center}
\vspace{-4.9cm}
\hspace{0.cm}
\begin{minipage}{16.cm}
\caption
{Experimental set-up. The trigger and trigger distances are not to scale}
\end{minipage}
\end{figure}
%--------------------fig.2------------------------
\begin{figure}
\begin{center}
\epsfig{file=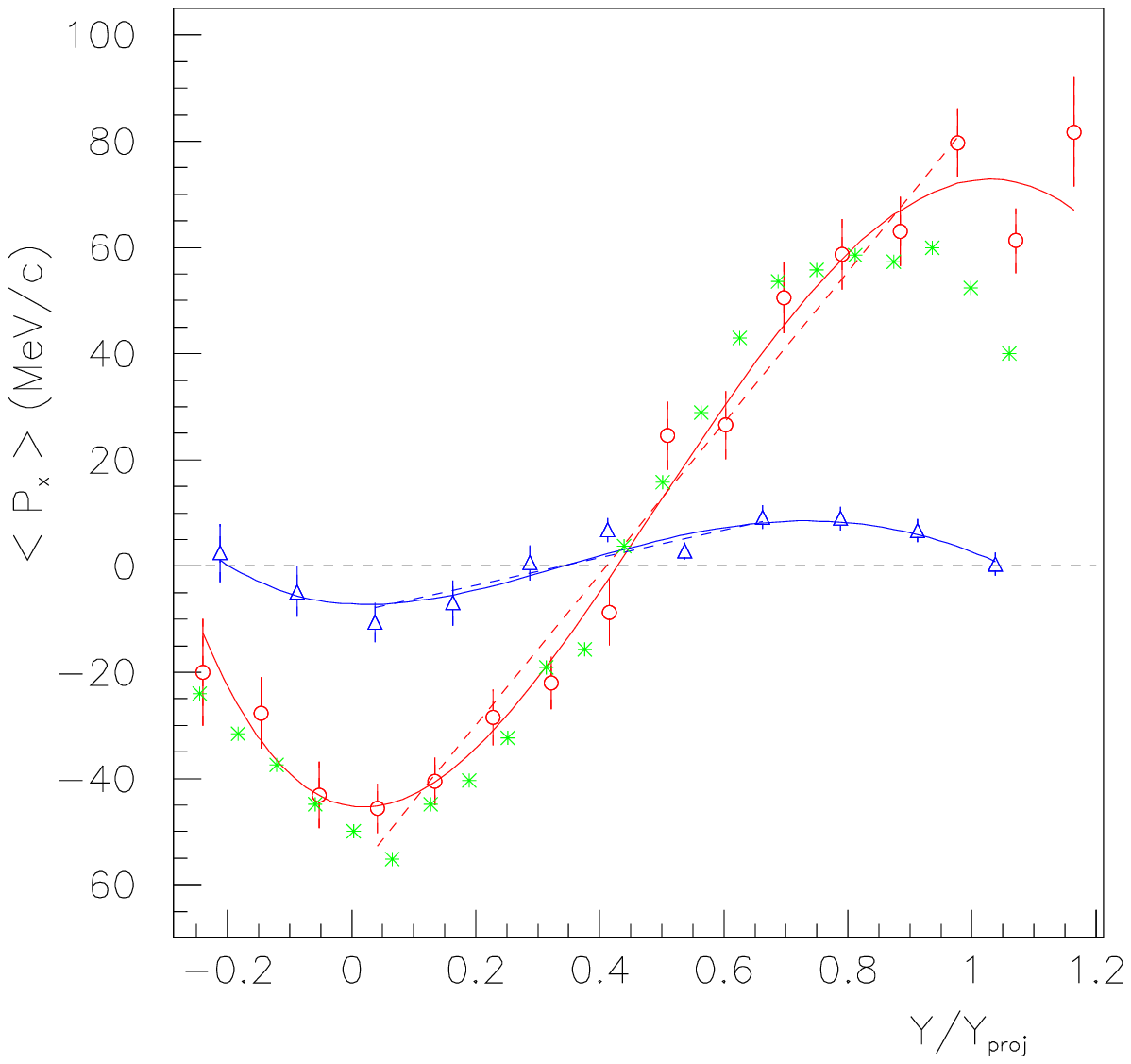,bbllx=0pt,bblly=0pt,bburx=594pt,bbury=842pt,
width=18cm,angle=0}
\end{center}
\vspace{-7.9cm}
\hspace{0.cm}
\begin{minipage}{16.cm}
\caption
{The dependence of $< P_{x} >$ on the normalized rapidity
 $y/y_{p}$ in the laboratory system
 in C-Ne  collisions.
{\Large\red $\circ$} -- for protons,
{\bl $\bigtriangleup$} -- for $\pi^{-}$ mesons,
{\Large\gre $\ast$} --  the QGSM data for protons.
The
lines represent linear fits of
experimental data in the intervals
 $0.01\leq y/y_{p} \leq0.90$ for protons
and $0.04\leq y/y_{p} \leq0.70$  for
$\pi^{-}$ mesons.
 The curves show
the 4th order polynomial fits.}
\end{minipage}
\end{figure}
%--------------------fig.3------------------------
\begin{figure}
\begin{center}
\epsfig{file=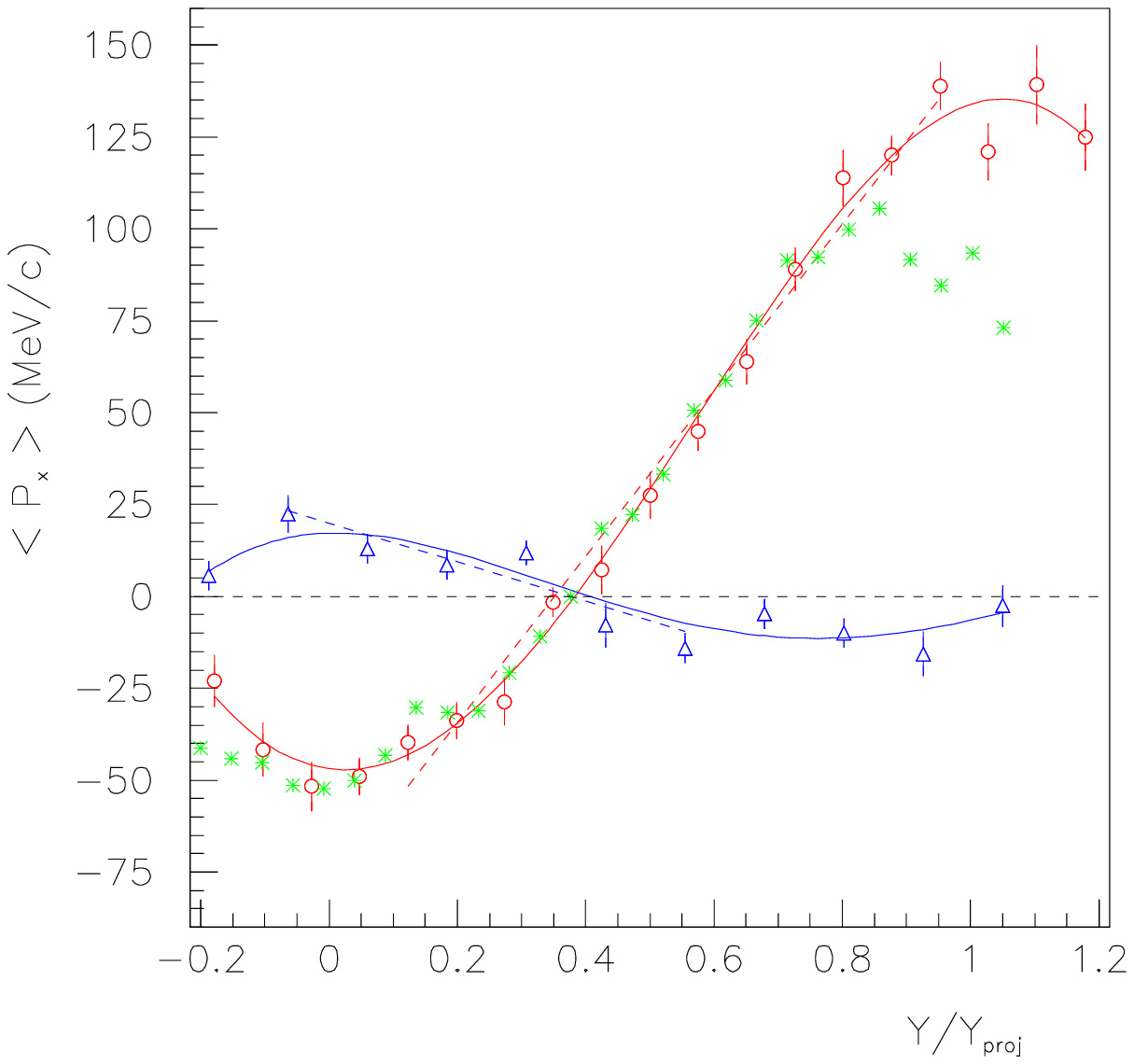,bbllx=0pt,bblly=0pt,bburx=594pt,bbury=842pt,
width=18cm,angle=0}
\end{center}
\vspace{-7.9cm}
\hspace{0.cm}
\begin{minipage}{16.cm}
\caption
{The dependence of $< P_{x} >$ on the normalized rapidity
 $y/y_{p}$ in the laboratory system
 in  C-Cu  collisions.
{\Large\red $\circ$} -- for protons,
{\bl $\bigtriangleup$} -- for $\pi^{-}$ mesons,
{\Large\gre $\ast$} --  the QGSM data for protons.
The
lines represent  linear fits of
experimental data in the intervals
 $0.01\leq y/y_{p} \leq0.90$ for protons
and $-0.06\leq y/y_{p} \leq0.6$  for
$\pi^{-}$ mesons.
 The curves show the
 4th order polynomial fits.}
\end{minipage}
\end{figure}
%--------------------fig.4 slac version------------------------
\pagebreak
\begin{figure}
\begin{center}
\epsfig{file=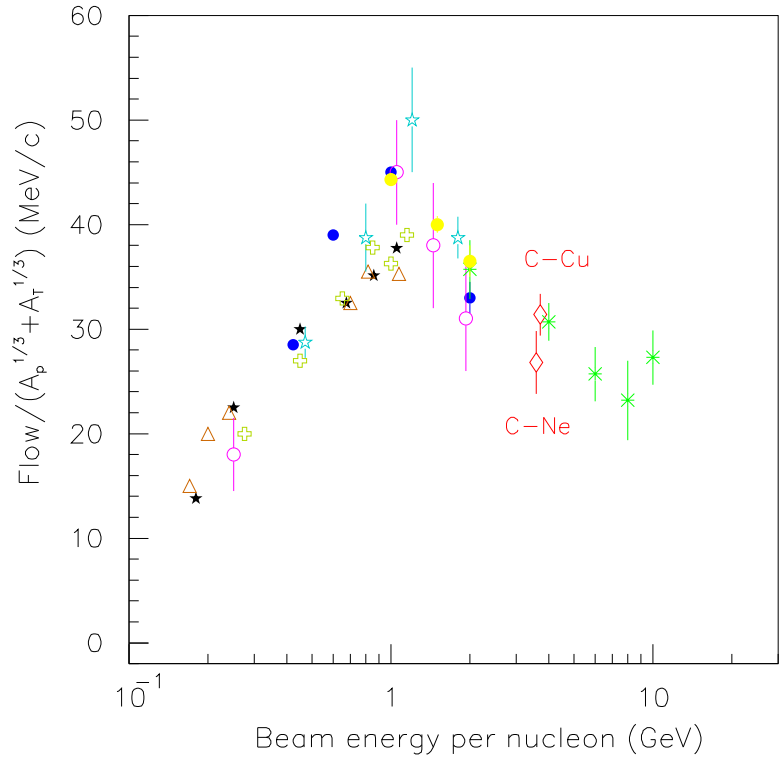,bbllx=0pt,bblly=0pt,bburx=594pt,bbury=842pt,
width=18.0cm,angle=0}
\end{center}
\vspace{-6.4cm}
\hspace{0.cm}
\begin{minipage}{16.0cm}
\caption
{Scaled flow values versus beam energy per nucleon, for different
projectile/target systems:
{\bla$\star$} -- Nb-Nb Plastic Ball,
{\bl$\bullet$}
-- Ni-Au EOS,
 {\br$\bigtriangleup$} -- Au-Au Plastic Ball,
{\Large\ma $\circ$} -- Ni-Ni FOPI,
{\ye$\bullet$} -- Ni-Cu EOS,{\Large\color{SpringGreen}\ +} -- Au-Au EOS,
{\Large\cy$\star$} -- Ar-Pb Streamer Chamber, the value at E=1.08 AGeV
represents
Ar-KCl
Streamer Chamber,
{\Large\red $\diamond$} -- C-Ne, C-Cu ( our result),{\Large\gre $\ast$} --
Au-Au E-895,
the value at
E=10 AGeV represents Au-Au from E-877.
%$\times$ -- Pb-Pb NA49.
 To improve the distinction between data
points at the same beam energy, some of the beam energy values
have been shifted.}
\end{minipage}
\end{figure}
%--------------------fig.5------------------------
%\pagebreak
\pagebreak
\begin{figure}
\begin{center}
\epsfig{file=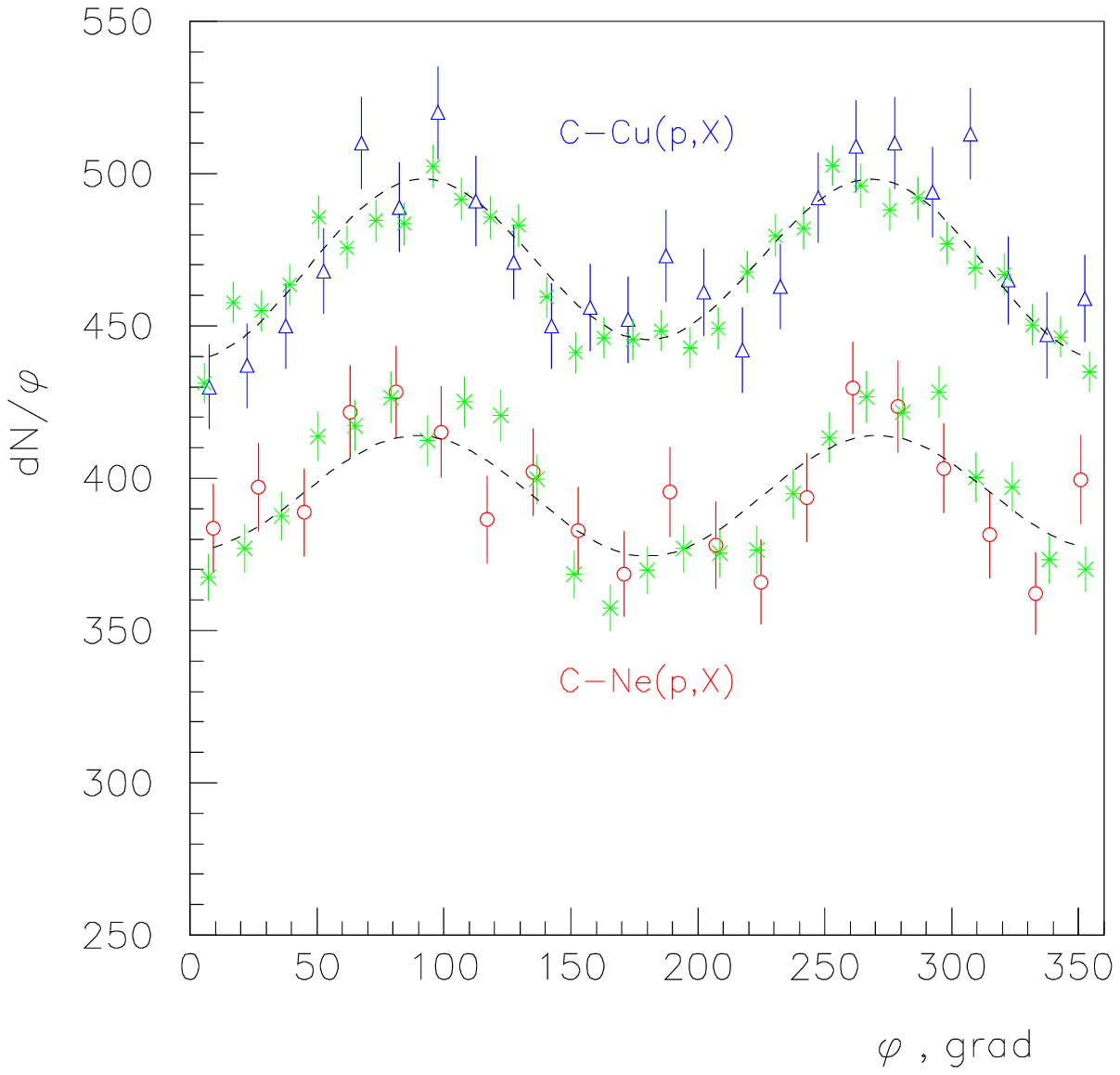,bbllx=0pt,bblly=0pt,bburx=594pt,bbury=842pt,
width=18cm,angle=0}
\end{center}
\vspace{-6.4cm}
\hspace{0.cm}
\begin{minipage}{16.0cm}
\caption
{The azimuthal distributions with respect to the reaction plane of midrapidity
protons dN/d$\phi$ .
{\Large\red $\circ$} -- for C-Ne ($-1\leq y_{cm}\leq1$),
{\bl $\bigtriangleup$} -- for C-Cu ($-1\leq y_{cm}\leq1$) interactions,
{\Large\gre $\ast$} --  the QGSM data.
Also shown are the fits using the function
{\bl$dN/d\phi=a_{0}(1+a_{1}cos\phi+a_{2}cos2\phi)$}.}
\end{minipage}
\end{figure}
%--------------------fig.6------------------------
%\pagebreak
\pagebreak
\begin{figure}
\begin{center}
\epsfig{file=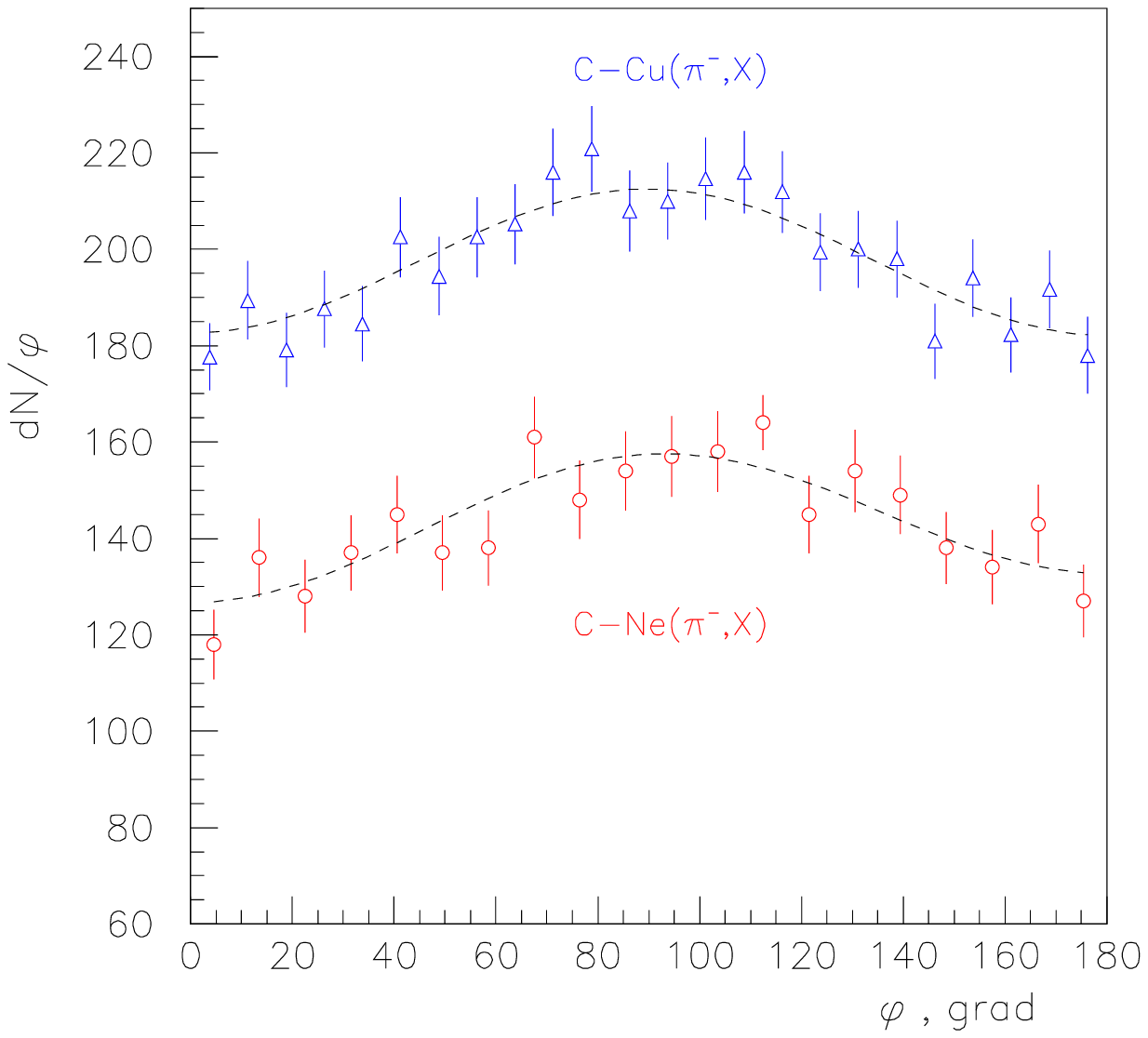,bbllx=0pt,bblly=0pt,bburx=594pt,bbury=842pt,
width=18cm,angle=0}
\end{center}
\vspace{-6.4cm}
\hspace{0.cm}
\begin{minipage}{16.0cm}
\caption
{The azimuthal distributions with respect to the reaction plane of midrapidity
$\pi^{-}$ mesons .
{\Large\red $\circ$} -- for C-Ne ($-1\leq y_{cm}\leq1$),
{\bl $\bigtriangleup$} -- for C-Cu ($-1\leq y_{cm}\leq1$) interactions,
{\Large\gre $\ast$} --  the QGSM data.
Also shown are the fits using the function
{\bl$dN/d\phi=a_{0}(1+a_{1}cos\phi+a_{2}cos2\phi)$}.}
\end{minipage}
\end{figure}
%--------------------fig.7 slac version------------------------
%\end{document}
\pagebreak
\begin{figure}
\begin{center}
\epsfig{file=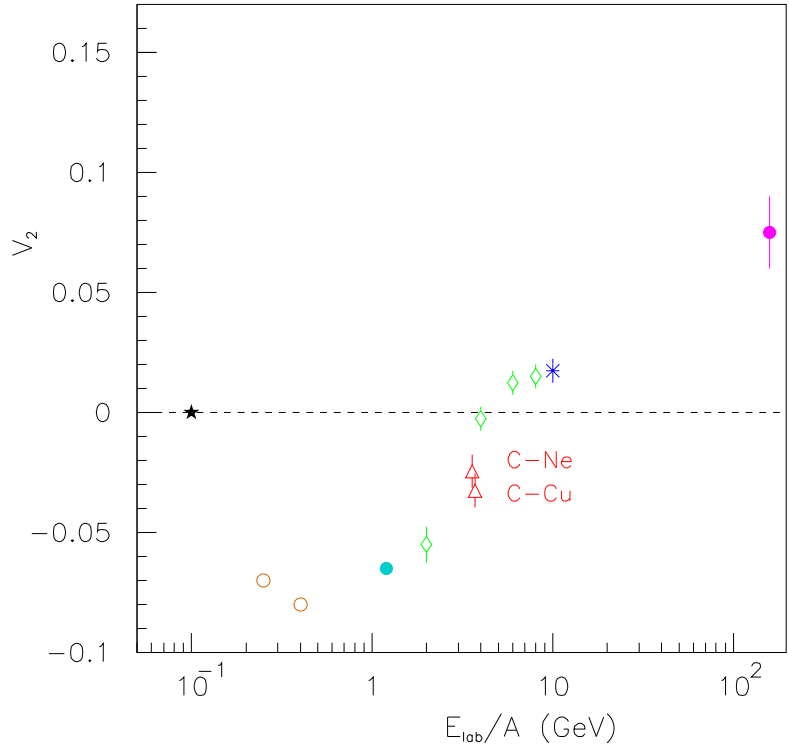,bbllx=0pt,bblly=0pt,bburx=594pt,bbury=842pt,
width=18cm,angle=0}
\end{center}
\vspace{-6.4cm}
\hspace{0.cm}
\begin{minipage}{16.0cm}
\caption
{The dependence of the Elliptic flow excitation function v$_{2}$ on energy
E$_{lab}$/A (GeV):
{\Large\bla$\star$} -- FOPI,{\Large\br $\circ$} -- MINIBALL,
{\cy $\bullet$} -- EOS,
{\gre$\diamond$}
%$\framebox(5,5){}}$
-- E-895,
{\Large\bl$\ast$} -- E-877,
{\ma$\bullet$}
-- NA49,
{\red$\bigtriangleup$} -- C-Ne, C-Cu (our results).}
\end{minipage}
\end{figure}
\end{document}